\documentclass[twocolumn,showpacs,preprintnumbers,amsmath,amssymb]{revtex4}
\usepackage{epsfig}
\usepackage{amsmath}

\begin{document}

\title{Knowledge Acquisition by Networks of Interacting Agents \\
in the Presence of Observation Errors}

\author{J. B. Batista and L. da F. Costa}
\affiliation{Institute of Physics at S\~ao Carlos, University of
S\~ao Paulo, P.O. Box 369, S\~ao Carlos, S\~ao Paulo, 13560-970
Brazil}

\date{24th April 2009}

\begin{abstract}
In this work we investigate knowledge acquisition as performed by
multiple agents interacting as they infer, under the presence of
observation errors, respective models of a complex system. We focus
the specific case in which, at each time step, each agent takes into
account its current observation as well as the average of the models
of its neighbors.  The agents are connected by a network of
interaction of Erd\H{o}s-R\'{e}nyi or Barab\'{a}si-Albert type. First
we investigate situations in which one of the agents has a different
probability of observation error (higher or lower). It is shown that
the influence of this special agent over the quality of the models
inferred by the rest of the network can be substantial, varying
linearly with the respective degree of the agent with different
estimation error. In case the degree of this agent is taken as a
respective fitness parameter, the effect of the different estimation
error is even more pronounced, becoming superlinear. To complement our
analysis, we provide the analytical solution of the overall behavior
of the system. We also investigate the knowledge acquisition dynamic
when the agents are grouped into communities. We verify that the
inclusion of edges between agents (within a community) having higher
probability of observation error promotes the loss of quality in the 
estimation of the agents in the other communities.
\end{abstract}

\maketitle

\vspace{0.5cm}
\emph{`Knowledge is of two kinds: we know a subject ourselves, or 
we know where we can find information upon it.' (S. Johson)}

\section{Introduction} 

Several important systems in nature, from the brain to society, are
characterized by intricate organization. Being naturally related to
such systems, humans have been trying to understand them through the
construction of models which can reasonably reproduce and predict the
respectively observed properties. Model building is the key component
in the scientific method. The development of a model involves the
observation and measurement of the phenomenon of interest, its
representation in mathematical terms, followed by simulations and
respective confrontation with further experimental evidences. Because
of the challenging complexity of the remaining problems in science,
model building has become intrinsically dependent on collaboration
between scientists or \emph{agents}. The problem of multiple-agent
knowledge acquisition and processing has been treated in the
literature (e.g. ~\cite{Botti:1993, Sajja:2008}), but often under
assumption of simple schemes of interactions between the agents
(e.g. lattice or pool).  Introduced recently, complex networks
(~\cite{Albert_Barab:2002,Dorogov_Mendes:2002, Newman:2003,
Boccaletti:2006, Costa_surv:2007}) have quickly become a key research
area mainly because of the generality of this approach to represent
virtually any discrete system, allied to the possibilities of relating
network topology and dynamics. As such, complex networks stand out as
being a fundamental resource for complementing and enhancing the
scientific method.

\begin{figure*}[htb]
  \vspace{0.3cm} 
  \begin{center}
  \includegraphics[width=0.7\linewidth]{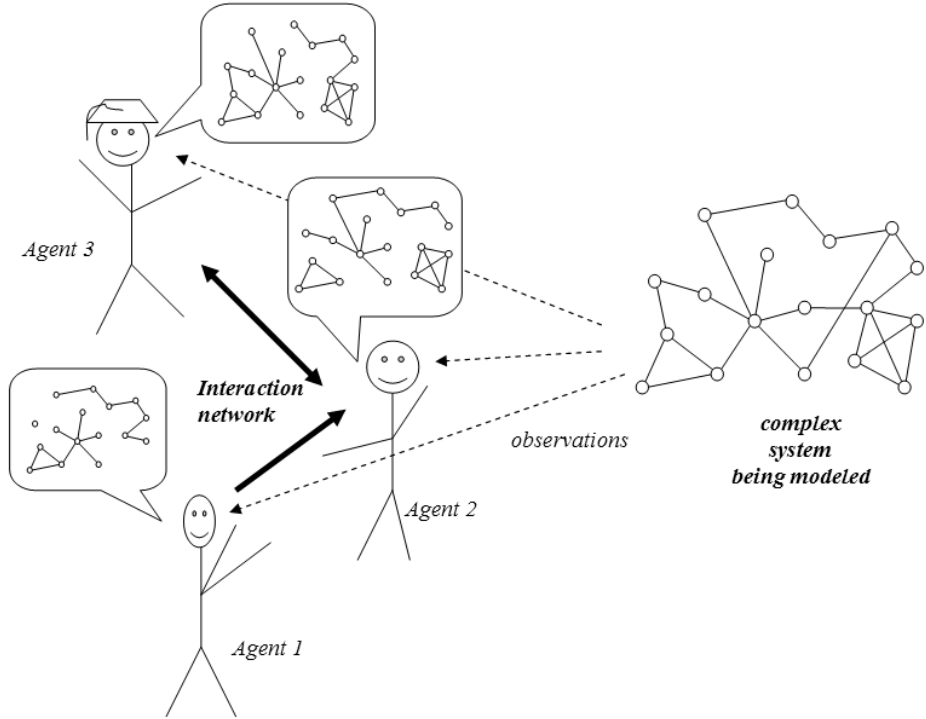} \\
  \vspace{0.5cm}
  \caption{Agents (scientists) develop their models of
           complex systems through observation and
           interactions.}~\label{fig:diagram} 
	  \end{center}
\end{figure*}

The present study addresses the issue of modeling how one or more
agents (e.g. scientists) progress while modeling a complex system. We
start by considering a single agent and then proceed to more general
situations involving several agents interacting through networks of
relationships (see Figure~\ref{fig:diagram}). The agents investigating
the system (one or more) are allowed to make observations and take
measurements of the system as they develop and complement their
respective individual models. Errors, incompleteness, noise and
forgetting are typically involved during a such model estimation. The
main features of interest include the quality of the obtained models
and the respective amount of time required for their estimation.  The
plural in `models' stands for the fact that the models obtained
respectively by each agent are not necessarily identical and will
often imply in substantial diversity. Though corresponding to a
largely simplified version of real scientific investigation, our
approach captures some of the main elements characterizing the
involvement of a large number of interacting scientists who
continuously exchange information and modify their respective models
and modeling approaches. As a matter of fact, in some cases the
development of models may even affect the system being modeled
(e.g. the perturbation implied by the measurements on the analyzed
systems).

Because interactions between scientists can be effectively represented
in terms of complex networks (e.g.~\cite{Price65:Science, Newman01_c,
Newman01_a, Newman01_b, Moody04, Barabasi02, silva2007cca}), it is
natural to resource to such an approach in our investigation.  It is
interesting to observe that the agents may not be limited to
scientists, but can also include intelligent machines and even
reference databases and libraries.  Though most of the previous
approaches to modeling scientific interaction in terms of complex
networks have focused on the topology of the collaborations, fewer
strategies (e.g.~\cite{Costa_know:2006}) have addressed the important
issue of how the dynamics of learning/knowledge acquisition evolves in
such systems, especially with respect to distinct patterns of
connectivity between the agents.  This is the main motivation of the
current work.

This article starts by presenting the general assumptions and
specifying and discussing successively more sophisticate levels of
modeling.  We then focus on the development of a model of a complex
system by a team of agents interacting through networks of
Erd\H{o}s-R\'{e}nyi (ER) and Barab\'{a}si-Albert (BA) types.  During
estimation, each agent takes into account not only its observation
(subject to a probability of error), but also the average of the
models of its neighbors. In order to quantify the performance of an
agent, we suggest an individual error based on the difference between
the model developed and the system under analysis.  Because the
influence of an agent is directly related to its degree, the overall
error of the system is expressed in terms of the mean of the
individual errors weighted by the degree of each respective agent.
The obtained results imply a series of interesting and important
insights, such as the identification of the substantial role of hubs
in affecting the estimation of models by all agents: while a linear
relationship is identified between the overall estimation errors and
the degree of single agents with different error rates, this
relationship becomes superlinear when the degree of each node is
considered as the respective fitness. In addition, in networks
characterized by presence of communities, intensifying the
interactions between agents having higher estimation errors in one of
the communities undermines the performance of the agents in the other
communities.

\section{General Assumptions}

The first important assumption is that the complex systems under study
be \emph{representable as a discrete structure} (e.g.\ a graph or
complex network), and that information about the several parts of this
system can be observed, often with a given probability of error.  By
\emph{agent}, it is meant any entity capable of making
observations/measurements of the system under analysis and storing
this information for some period of time.  Therefore, any scientist
can be naturally represented as an agent. However, automated systems,
from measurement stations to more sophisticated reasoning systems, can
also be represented as agents. Actually, even less dynamic systems
such as books or libraries can be thought of as a sort of passive
agents, in the sense that they evolve (new editions and versions) as a
consequence of incorporation of the evolution of knowledge.  

Each agent is capable of making \emph{observations/measurements} of
the system under investigation. The process of measurement typically
involves errors, which can be of several types such as observing a
connection where there is none. Such errors can be a direct
consequence of the limited accuracy of measurements devices as well as
of the priorities and eventual biases (not to mention wishful
thinking) of each agent. Several other possibilities, such as the
existence of non-observable portions of the system, can also be
considered and incorporated into the model. In addition, the model
kept by each agent may undergo degradation as a consequence of noise
and/or vanishing memory.

A particularly important element in our approach is the incorporation
of different types of interactions between the agents, which can be
represented in terms of a graph or network (see
Figure~\ref{fig:diagram}). In this case, each agent gives rise to a
node, while interactions (e.g.\ collaborations) among them are
represented by links. The single-agent configuration can be
immediately thought of as a special case where the graph involves only
one node (agent).  Several types of interactions are possible
including, but being not limited to conversations, attendance to talks
or courses (the speaker becomes a temporary hub), article/book,
reading, and Internet exchanges (e.g. e-mailing and surfing). Such
interactions may also involve errors (e.g. misunderstanding words
during a talk), which can be eventually incorporated into our
model. It is interesting to observe that the network of interaction
between the agents may present dynamic topology, varying with time as
a consequence of new scientific partnerships, addition or removal of
agents, and so on.  

Therefore, the framework considered in our investigation includes the
following three basic components: (i) the complex system under
analysis $S$; (ii) one or more agents capable of taking
observations/measurements of the system $S$ (subject to error) and of
interacting one another; (iii) a network of interaction between the
agents.  In the following we address, in progressive levels of
sophistication, the modeling of knowledge acquisition/model building
in terms of complex networks concepts and tools.

\section{Single-Agent Modeling}

A very simple situation would be the case where a single agent is
allowed to observe, in uniformly random fashion, the presence or not
of connections between all pairs of nodes of the complex system being
modeled. The observation of each edge involves a probability $\gamma$
of error, i.e. observing a connection where there is none and
vice-versa.  A possible procedure of model building adopted by the
agent involves taking the average of all individual observations up to
the current time step $T$ , i.e.

\begin{equation}
  \left< x \right>_T = \frac{1}{T} \sum_{t=1}^{T} x_t,
\end{equation}

where $x$ is the value of a specific edge (0 if non-existent and 1
otherwise)and $x_t$ is the observation of $x$ at time step
$t$. Observe that we are considering the observation error to be
independent along the whole system under analysis. Let us quantify the
error for estimation of any edge, after $T$ time steps as follows

\begin{equation}
  \epsilon(x)_T = \left| x_* - \left< x \right>_T \right|,
\end{equation}

where $x_*$ is the original value of that edge.  It can be easily
shown that

\begin{equation}
  \lim_{T \rightarrow \infty}\epsilon(x)_T = \gamma
\end{equation}

Because the observation error is independent among the pairs of nodes,
the average of the errors along the whole network is identical to the
limit above.

Thus, in this configuration the best situation which can be aimed at
by the agent is to reach a representation of the connectivity of the
original complex system up to an overall limiting error $\gamma$,
reached after a considerable period of time.  Though the speed of
convergence is an interesting additional aspect to be investigated
while modeling multi-agent knowledge acquisition, we leave this
development for a forthcoming investigation.

\section{Multiple-Agent Modeling}

We now turn our attention to a more interesting configuration in which
a total of $N_a$ agents interact while making observations of the
complex system under analysis, along a sequence of time steps. As
before, each agent observes the whole system at each time step, with
error probability $\gamma$.  In case no communication is available
between the agents, each one will evolve exactly as discussed in the
previous section. However, our main interest in this work is to
investigate how \emph{interactions} and \emph{influences} between the
multiple agents can affect the quality and speed at which the system
of interest is learned. The original system under study is henceforth
represented in terms of its respective adjacency matrix $A_*$. We also
assume that the agents exchange information through a complex network
of contacts.

One of the simplest, and still interesting, modeling strategies to be
adopted by the agents of such a system involves the following
dynamics: at each time step $t$, each of the agents $i$ observes the
connectivity of the complex system with error $\gamma$, yielding the
adjacency matrix $O_i^t$, and also receives the current matrices from
each of its immediate neighbors$ j$ (i.e. the agents to which it is
directly connected). The agent $i$ then calculates the mean matrix
$\left< K \right>^t$ of the matrices $K_{ji}^t$ received from all its
neighbors $j$, i.e.

\begin{equation}~\label{eq:ng_av}
  \left< K \right>^t = \frac{1}{k_i}\sum_{j} K_{ji}^t,  
\end{equation}

where $k_i$ is the degree of the agent $i$. The agent $i$ then makes a
weighted average between its current matrix $K_i^t$ and the immediate
neighbors mean matrix, i.e.

\begin{equation}
  v_i^{t} = (a \left< K \right>^t + (1-a)K_i^t),
\end{equation}

where $0 < a \leq 1$ is a relative weight.  We henceforth assume $a =
0.5$, so that $v_i^{t}$ becomes equal to the average between its
current matrix and of the neighbors at that time step. Each agent $i$
subsequently adds this value to its current observation, i.e.

\begin{equation}
  K_i^{(t+1)} = v_i^{(t)} + O_i^{(t)},
\end{equation}

so after $T$ time steps the estimation of the adjacency matrix by the
agent $i$ can be given as

\begin{equation}
  A_i^{(T)} = K_i^{(T)} / T.
\end{equation}

This simple discrete-type dynamics has some immediate important
consequences.  Because each node receives estimations from its
neighbors at each time step, an altered quality of estimation is
obtained. Indeed, by averaging between its current observation and the
mean estimation from the neighbors, even the limit error may be
actually modified with respect to the single agent situation.

Several variations and elaborations of this model are possible,
including the consideration of noise while transmitting the
observations between adjacent agents, forgetting, the adoption of
other values of $a$, as well as other averaging and noise schemes. In
this article, however, we focus attention on the multi-agent model
described in this section with error being present only during the
observation by each agent. In the remainder of our work, we report
results of numerical simulation considering three particularly
important situations: (i) multiple-agents with equal observation
errors; (ii) as in (i) but with one of the agents having different
observation error; and (iii) multiple-agents scheme applied in
networks with community structure and one of the communities having
higher error rate. Interesting results are obtained for all these
three configurations including the analytical solution of the overall
behavior of the system.

\section{Case Examples: Random and Scale-Free Networks of Agents}

In this section we investigate further the dynamics of multi-agent
learning by considering theoretical simulations performed with a fixed
collaboration network. We assume that the agents collaborate through a
uniformly random model (ER) as well as a scale-free model (BA), both
containing 50 nodes and average degree equal to 9.4. For simplicity's
sake, we consider only a single realization of such a model in our
subsequent investigation.  The performance of each agent is quantified
in terms of the error between the original network and the models
obtained by that agent after a sufficiently large number of time steps
(henceforth assumed to be equal to $T = 300$ time steps).  This error
is calculated as

\begin{equation}
  \varepsilon_{p}^{(T)} = \frac{1}{N^2} \sum_{m=1}^{N}\sum_{n=1}^{N} |A_i^{(T,m,n)} - A_{*}^{m,n}|
\end{equation}

where $A_i$ is the adjacency matrix representation of the model of
agent $i$ at time step $T$. The original network to be learnt by the
agents is a Barab\'{a}si-Albert network containing $N=50$ nodes and
average degree equal to 6, represented by the respective adjacency
matrix $A_*$. The overall performance of the network, including all
its agents, is henceforth expressed in terms of the average of the
above error weighted by the degree considering all agents, which is
here called the \emph{overall error} of the system:

\begin{equation}
  E^{(T)} = \frac{1}{\sum_{i=1}^{N_a} k_i} \sum_{i=1}^{N_a} k_i\varepsilon_{i}^T
\end{equation}

Because the influence of an agent in the network is correlated with
the number of its neighbors, such a definition of the overall error
takes into account the respective importance of the models of each
agent.

\subsection{Analytical Description}~\label{analyitical}

We now present the mathematical formalization of the above concepts.
We start with the difference equation that characterizes the dynamics
of each agent $i$:

\begin{equation}
  2K_i^{t+1} = \left[\frac{1}{k_i}\sum_{j\in\Gamma}K_j^t\right] +K_i^t+2O_i^t
\end{equation}

where $K_i^t$ is the current matrix and $O_i^t$ is the observation of
the agent $i$ at time step $t$. The term in brackets is the mean of
the matrices received from the set $\Gamma$ of the neighbors of
$i$. Now we consider the continuous approximation given by:

\begin{equation}
  2\frac{d}{dt}K_i(t) = \left[\frac{1}{k_i}\sum_{j\in\Gamma}K_j(t)\right] -K_i(t)+2O_i(t)
\end{equation}

Dividing by $t$ and knowing that
$\frac{1}{t}\frac{d}{dt}K_i(t)=\frac{d}{dt}\left(\frac{K_i^t}{t}\right)+\frac{1}{t^2}K_i^t$, we have

\begin{multline}
2k_i\frac{d}{dt}\left[\frac{K_i(t)}{t}\right] + 2k_i\frac{K_i(t)}{t^2} =\\
=\left[\sum_{j\in\Gamma}\frac{K_j(t)}{t}\right] -k_i\frac{K_i(t)}{t}+2k_i\frac{O_i(t)}{t}
\end{multline}

and adding, for every agent $i$ :

\begin{multline}
  \frac{d}{dt}\sum_{i=1}^{N_a}\left[\frac{k_iK_i(t)}{t}\right] = \\
\frac{1}{t}\sum_{i=1}^{N_a}k_iO_i(t) - \frac{1}{t}\sum_{i=1}^{N_a}\left[\frac{k_iK_i(t)}{t}\right]
\end{multline}

since $\sum_i\sum_{j\in\Gamma}\frac{K_j(t)}{t}=\sum_{i}\frac{k_{i}K_i(t)}{t}$. Finally, we define  $J\equiv\sum_{i}\frac{k_{i}K_{i}(t)}{t}$:

\begin{equation}~\label{eq_14}
  \frac{d}{dt}J(t)= \frac{1}{t}\sum_{i=1}^{N}k_iO_i(t) - \frac{1}{t}J(t)
\end{equation}

For an agent $i$ the probability of observing an edge where there is
none and vice-versa is $\gamma_i$.  So, for an entry $A_*^{m,n}$ of
the system under observation:

\begin{multline}
  \lim_{T \rightarrow \infty}\left[\frac{1}{T}\int\left(\sum_{i}k_iO_i^{m,n}(t) \right) dt \right] = \\
=\left\{\begin{array}{lc}
\sum_{i}k_i(1-\gamma_{i})&\mbox{if}\quad A_*^{m,n}= 1\\ \noalign{\medskip}
\sum_{i}k_i(\gamma_{i})&\mbox{if}\quad A_*^{m,n}= 0
\end{array}\right.
\end{multline}

Applying this result in equation ~\ref{eq_14} and solving, we have :

\begin{equation}
  \lim_{T \rightarrow \infty}J^{m,n}(T) = 
\left\{\begin{array}{lc}
\sum_{i}k_i(1-\gamma_{i})&\mbox{if}\quad A_*^{m,n}= 1\\ \noalign{\medskip}
\sum_{i}k_i(\gamma_{i})&\mbox{if}\quad A_*^{m,n}= 0
\end{array}\right.
\end{equation}

and subtracting $\sum_{i}k_iA_*^{m,n}$ and adding for all m and n 

\begin{multline*}
  \sum_{m=1}^{N}\sum_{n=1}^{N}\left( J^{m,n}(T)-\sum_{i}k_{i}A_*^{m,n} \right) = \\
=\sum_{i} \left( \sum_{m=1}^{N} \sum_{n=1}^{N}\gamma_{i}k_i \right)
\end{multline*}

\begin{equation}
\Rightarrow
  \sum_{i=1}^{Na} \left[ \frac{k_i}{N^2} \left( \frac{K_i^{T} }{T}- A_* \right) \right] = \sum_{i=1}^{N_a}k_i\gamma_i
\end{equation}

for both cases. The term in the sum is precisely the individual error
of an agent $i$ weighted by its degree. Then, dividing by the sum of
the degrees, we have the analytical expression for the overall error
of the system.

\begin{equation*}
   \frac{\sum_{i=1}^{Na}k_i\varepsilon_i}{\sum_{i}k_i} = \frac{\sum_{i=1}^{Na}k_i\gamma_i}{\sum_{i}k_i}
\end{equation*}

\begin{equation}
\Rightarrow
   E^T = \frac{1}{\left< k\right> N_a}\sum_{i=1}^{N_a}k_i\gamma_i
\end{equation}

We can apply this result in two following situations, assuming a
network of agents with average degree $\left< k \right>$ and maximum
and minimum degree equal to $max_d$ and $min_d$, respectively:
\\\\
\emph{a) One agent $j$ with degree $k_j$ having an observation error $\gamma_j=p\gamma$ and all other agents with an error rate $\gamma$}:

\begin{equation*}
   E^T = \frac{\left( \sum_{i=1}^{N_a}{k_i\gamma} \right) -\gamma k_j
   + p\gamma k_j}{\left< k\right> N_a}
\end{equation*}

\begin{equation}
\Rightarrow
   E^T = \frac{k_j\gamma (p-1)}{\left< k\right> N_a}+\gamma
\end{equation}

which is a linear relation with respect to the degree of the agent
with different error rate. Note that, there is no difference between
any two networks of agents with the same average degree and number
of vertices.
\\\\
\emph{b) One agent $j$ with degree $k_j$ having an observation error 
proportional to its degree $\gamma_{j}(\gamma \ge\gamma_{j} \ge1)$ and
all other agents with an error rate $\gamma$:}

\begin{equation}
\Rightarrow
   E^T = \frac{k_j(1-\gamma) (k_j-min_d)}{\left< k\right> N_a (max_d-min_d)}+\gamma
\end{equation}

which is a power-law relation with respect to the degree of the agent
with different error rate.

\subsection{Fixed Observation Errors}~\label{eq_errs}

\begin{figure}[htb]
  \vspace{0.3cm} 
  \begin{center}
  \includegraphics[width=0.9\linewidth]{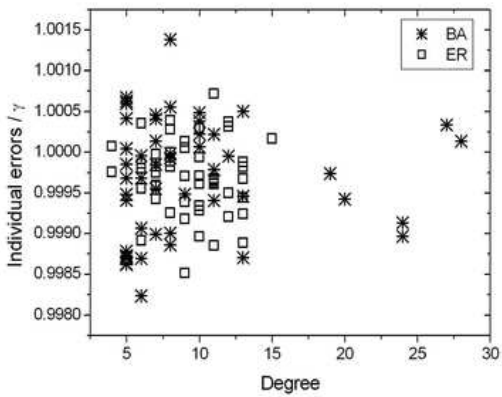} \\
  \vspace{0.5cm}
  \caption{The individual estimation errors in terms of the degree of
  the agents obtained for the Erd\H{o}s-R\'{e}nyi (ER) and
  Barab\'{a}si-Albert (BA) models with $N_a=50$ and $\left< k \right>
  =9.4$.}~\label{fig:eq_errors_nofit} \end{center}
\end{figure}

In this first configuration, all nodes have the same estimation error
$\gamma = 0.2$, and therefore the same influence, over the averages
obtained by each agent.  However, nodes with higher degree, especially
hubs, are still expected to influence more strongly the overall
dynamics as a consequence of the fact that their estimated models are
taken into account by a larger number of
neighbors. Figure~\ref{fig:eq_errors_nofit} shows the distribution of
the average of the errors $\varepsilon_i^T$ obtained for this case in
terms of the respective degrees. As could be expected, the errors
obtained among the agents are very similar one another. Thus, there is
no major difference in the learning quality among the agents. A rather
different situation arises in the next sections, where we consider
different error rates.

\subsection{Varying Observation Errors}~\label{sec:ve_nf}

We now repeat the previous configuration in two new situations: (i)
one of the agents having half or twice the error rate of the others;
and (ii) having an error proportional to its degree. Simulations are
performed independently while placing the higher error rate at all of
the possible nodes, while the respective overall errors are
calculated. In addition, we compared the behavior for the Erd\H{o}s-R\'{e}nyi
(ER) and Barab\'{a}si-Albert (BA) network of agents.

\begin{figure*}[htb]
  \vspace{0.3cm} 
  \begin{center}
  \includegraphics[width=0.45\linewidth]{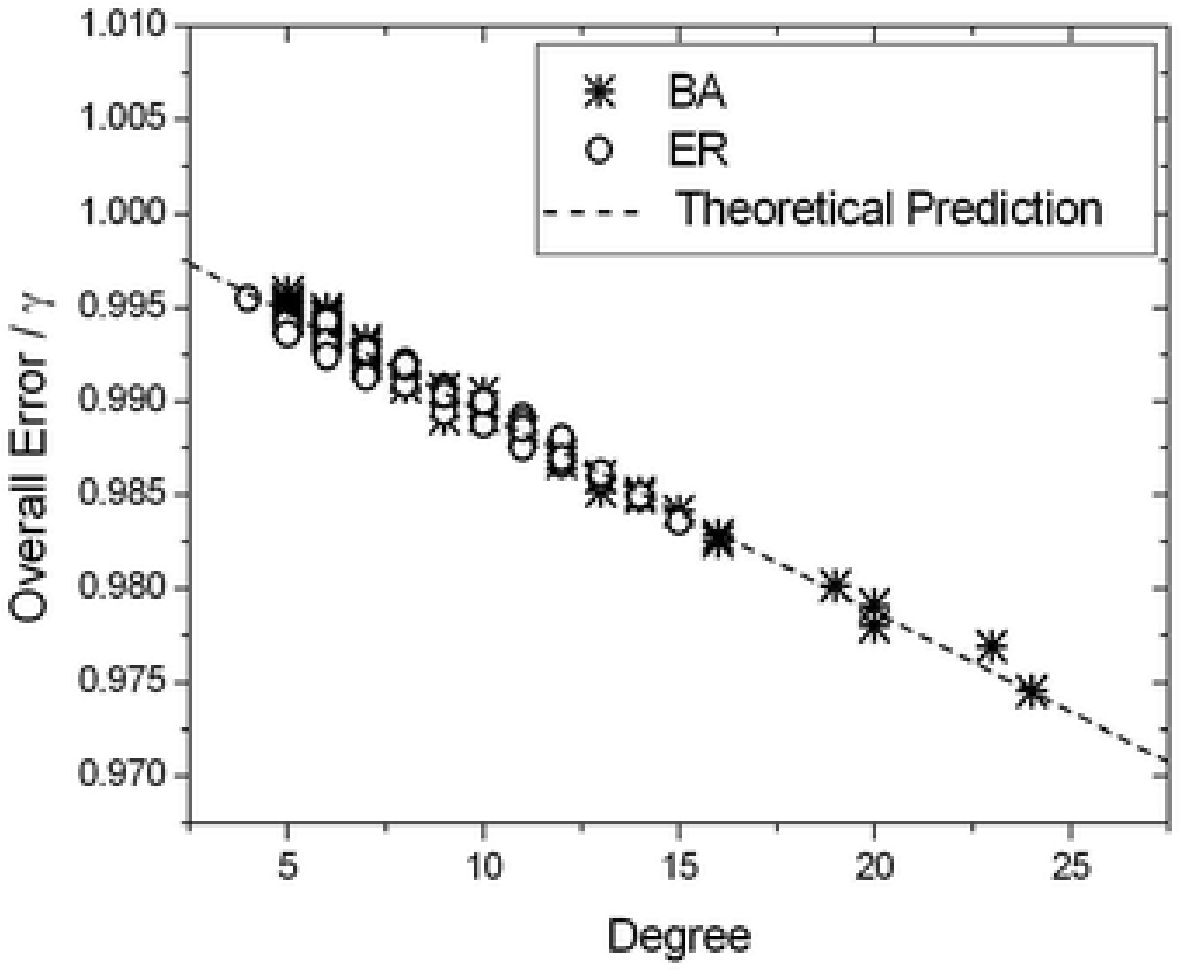} 
  \includegraphics[width=0.45\linewidth]{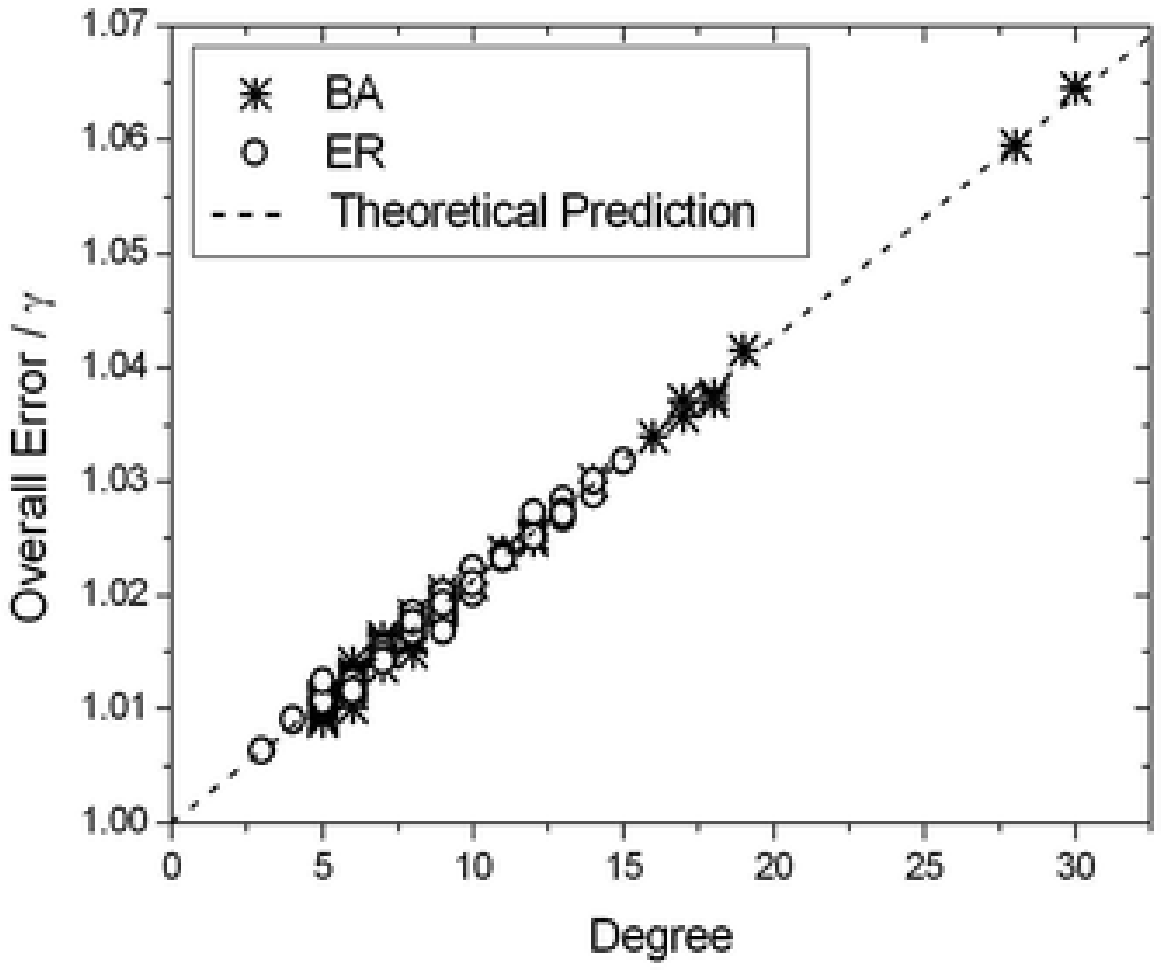} \\
   (a)   \hspace{8cm}  (b) \\
  \vspace{0.5cm}
  \caption{The overall estimation errors in terms of the degree of the
  agents with different error rates for ER and BA models: (a) one
  agent with an error rate equal to $0.2$ and all other agents with
  $\gamma = 0.4$; and (b) one node with 0.4 and all other nodes with
  $\gamma = 0.2$. The dashed line represents the analytical
  solution.}~\label{fig:dif_errors_nofit} \end{center}
\end{figure*}

Figure~\ref{fig:dif_errors_nofit}(a) shows the results obtained while
considering error probability equal to 0.2 for one of the agents and
$\gamma = 0.4$ for all other agents. It is possible to identify
a substantial difference of the mean quality of the models which depends
linearly on the degree of the node with different error rate. More
specifically, it is clear that the overall errors are much smaller for
nodes with higher degrees. In other words, the best models will be
obtained when the hubs have smaller observation errors, influencing
strongly the rest of the agents through the diffusion, along time, of their
respectively estimated models.

Figure~\ref{fig:dif_errors_nofit}(b) shows the results obtained when
the estimation error of one of the agents is $0.4$ while the rest of
the agents have error rate $\gamma = 0.2$.  The opposite effect is verified,
with the overall error increasing linearly with the degree of the node
with higher estimation error. Moreover, as expected, no difference was
found between the ER and BA models in both cases.

\begin{figure}[htb]
  \vspace{0.3cm} 
  \begin{center}
  \includegraphics[width=0.9\linewidth]{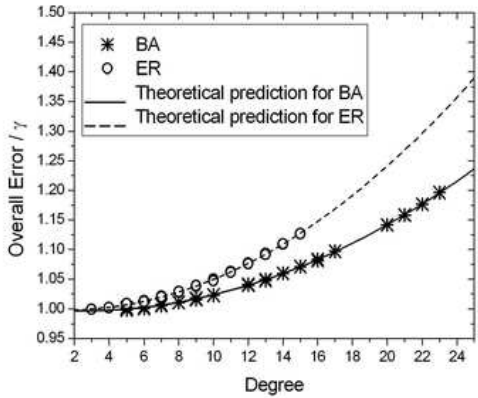} \\
  \vspace{0.5cm}
  \caption{The overall estimation errors in terms of the degree of the
  agents with higher error rates: one agent with an error proportional
  to its degree and all other agents with $\gamma = 0.2$. The dashed
  and straight lines represent the analytical solution for the ER and
  BA models, respectively. }~\label{fig:dif_errors_fit} \end{center}
\end{figure}

Figure~\ref{fig:dif_errors_fit} shows the results obtained when one of
the agents has a higher error rate than the others, and proportional
to its degree. Results similar to the previous case were found, but
with a nonlinear behavior. In addition, accordingly with the
analytical predictions, a separation was verified between the ER and
BA networks (ER presents higher values than BA). This means that when
we consider two nodes with the same degree, each in one network, the
overall error is lower for the BA model, since their hubs influence
more strongly the to increase the quality of the estimated models.

\begin{figure*}[htb]
  \vspace{0.3cm} \begin{center}
  \includegraphics[width=0.45\linewidth]{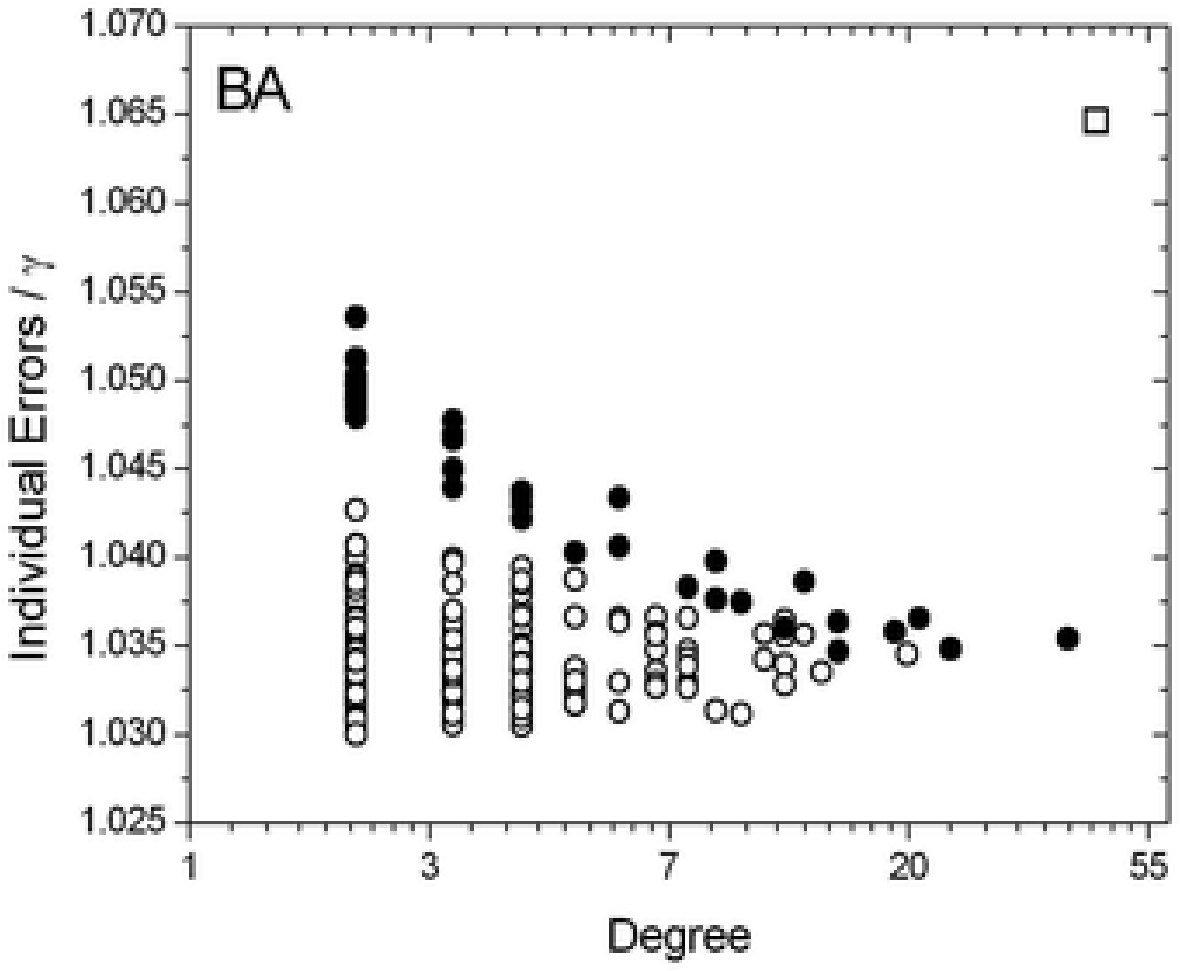}
  \includegraphics[width=0.45\linewidth]{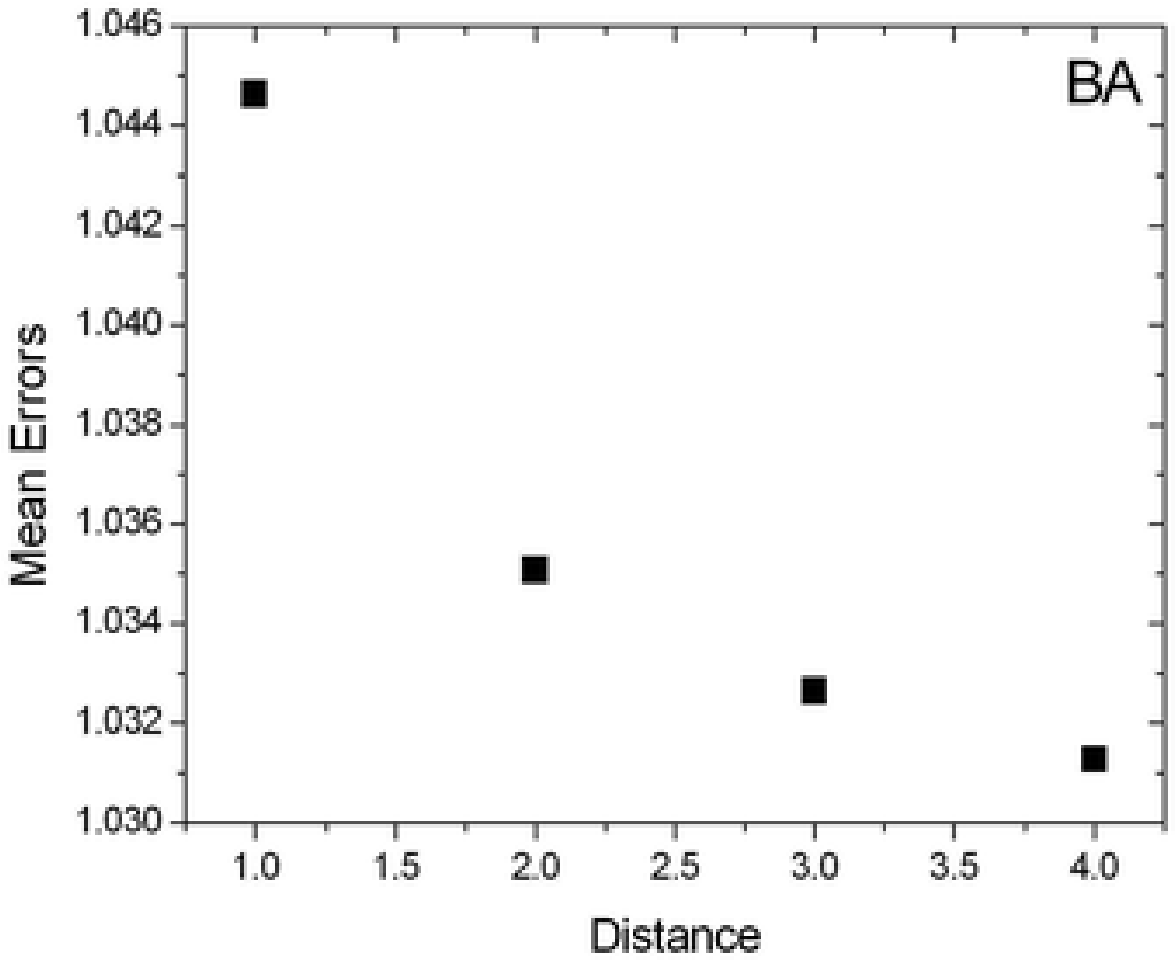} \\
  (a) \hspace{8cm} (b) \\ \vspace{0.5cm} 
  \includegraphics[width=0.45\linewidth]{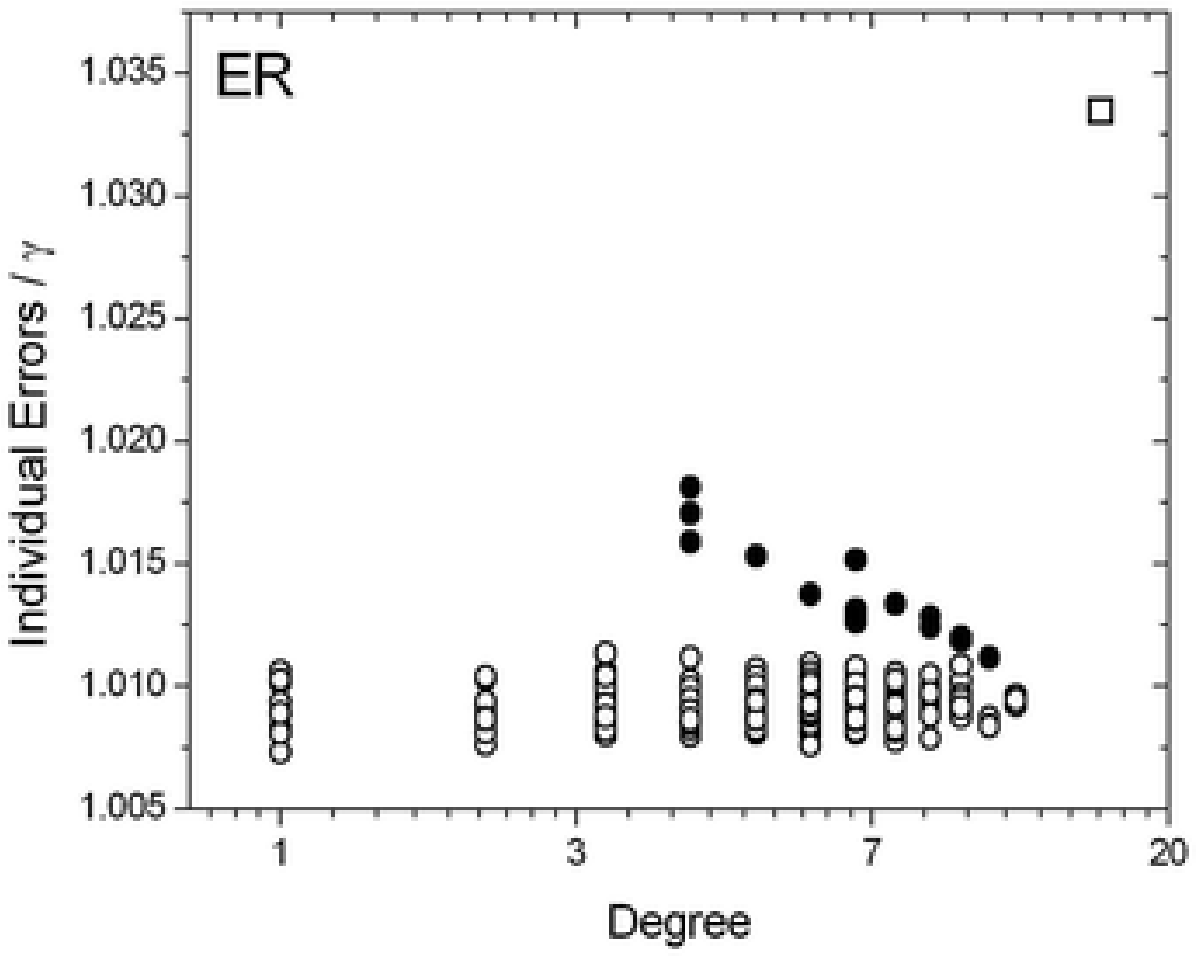}
  \includegraphics[width=0.45\linewidth]{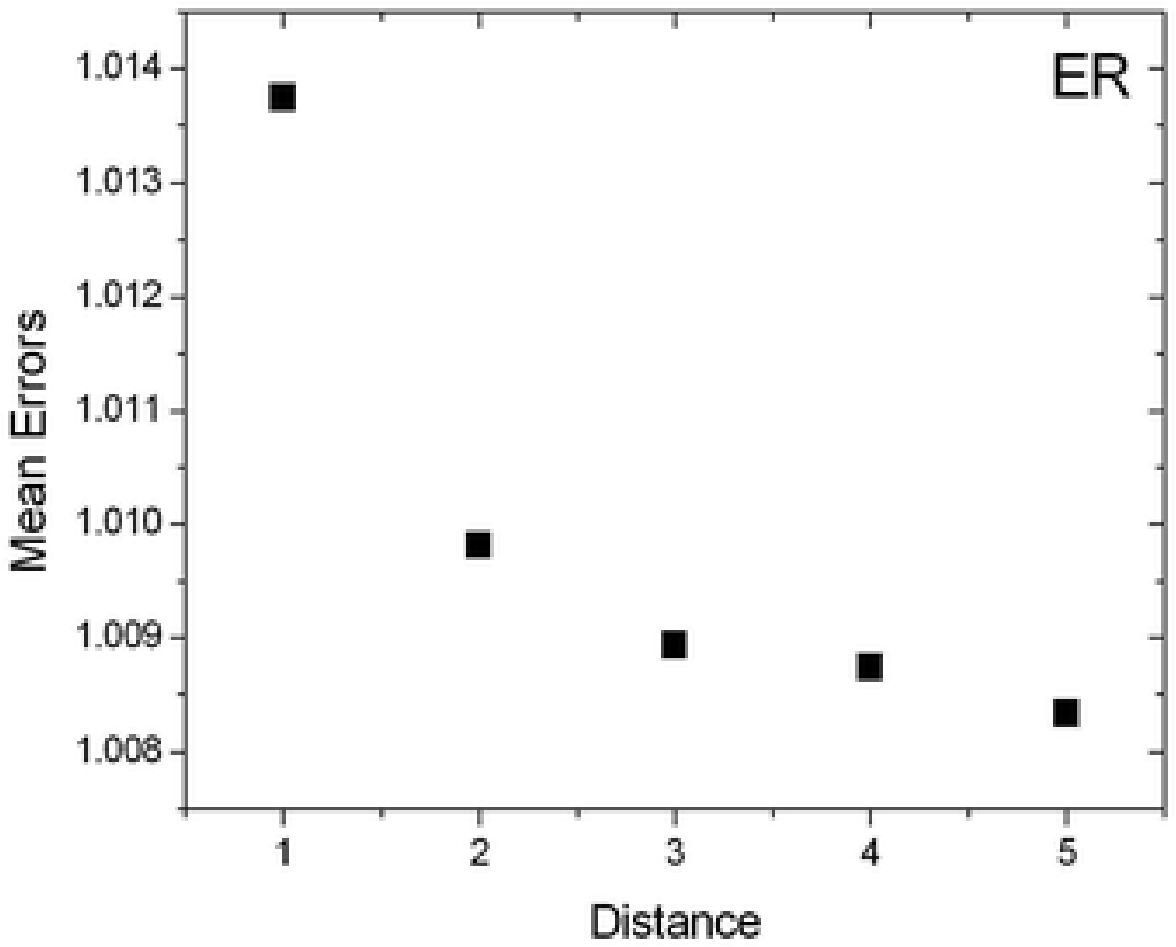} \\
  (c) \hspace{8cm} (d) \\ \vspace{0.5cm} 
  \caption{The individual estimation errors in terms of the degree of
  the agents (a and c): $\square$, the agent with highest error rate;
  $\bullet$, the neighborhood of the agent with different error
  rate. Figures (b and d) show the mean value of the individual
  errors of a set of nodes with the same topological distance from the agent
  with higher error rate. Simulations for the ER and BA models
  with $N_a=300$ and $\left< k \right> =4.7$.}~\label{fig:errs_deg} \end{center}
\end{figure*}

A better picture of the influence of the degree over the model
development by other agents is provided in Figure~\ref{fig:errs_deg}
respectively to the situation having twice the error probability
($\gamma_j = 0.4$) for the agent $j$ with the highest
degree. Figures~\ref{fig:errs_deg}(a) and ~\ref{fig:errs_deg}(c)
consider the $\varepsilon_i^T$ of individual agents in terms of their
respective degrees for the BA and ER models, respectively. It is clear
from these results that the degree of the differentiated agent affects
the estimation of the whole set of agents, especially for its
neighborhood, shifting substantially the average individual errors.
Observe also that, in both cases, the largest individual error results
precisely at the less accurate agent. Furthermore, nodes with high
degrees tend to be less influenced by the agent with different error
rate.

Figures ~\ref{fig:errs_deg}(b) and ~\ref{fig:errs_deg}(d) complement
the analysis considering the mean value of the individual errors of
agents with the same topological distance from the agent with
different error probability. The results show that the closer agents
tend to be more affected than the peripherals.

From the results above, we immediately verify that the presence of the
less accurate agent produces a non-uniform error distribution.  Since
the individual error of an agent is produced by averaging between its
model and the mean estimation from the neighbors, the direct contact
(and proximity) with agents with higher error rates alters considerably
its final model. Moreover, this effect is more strongly felt by agents with few
connections.

\begin{figure*}[htb]
  \vspace{0.3cm} 
  \begin{center}
  \includegraphics[width=0.3\linewidth]{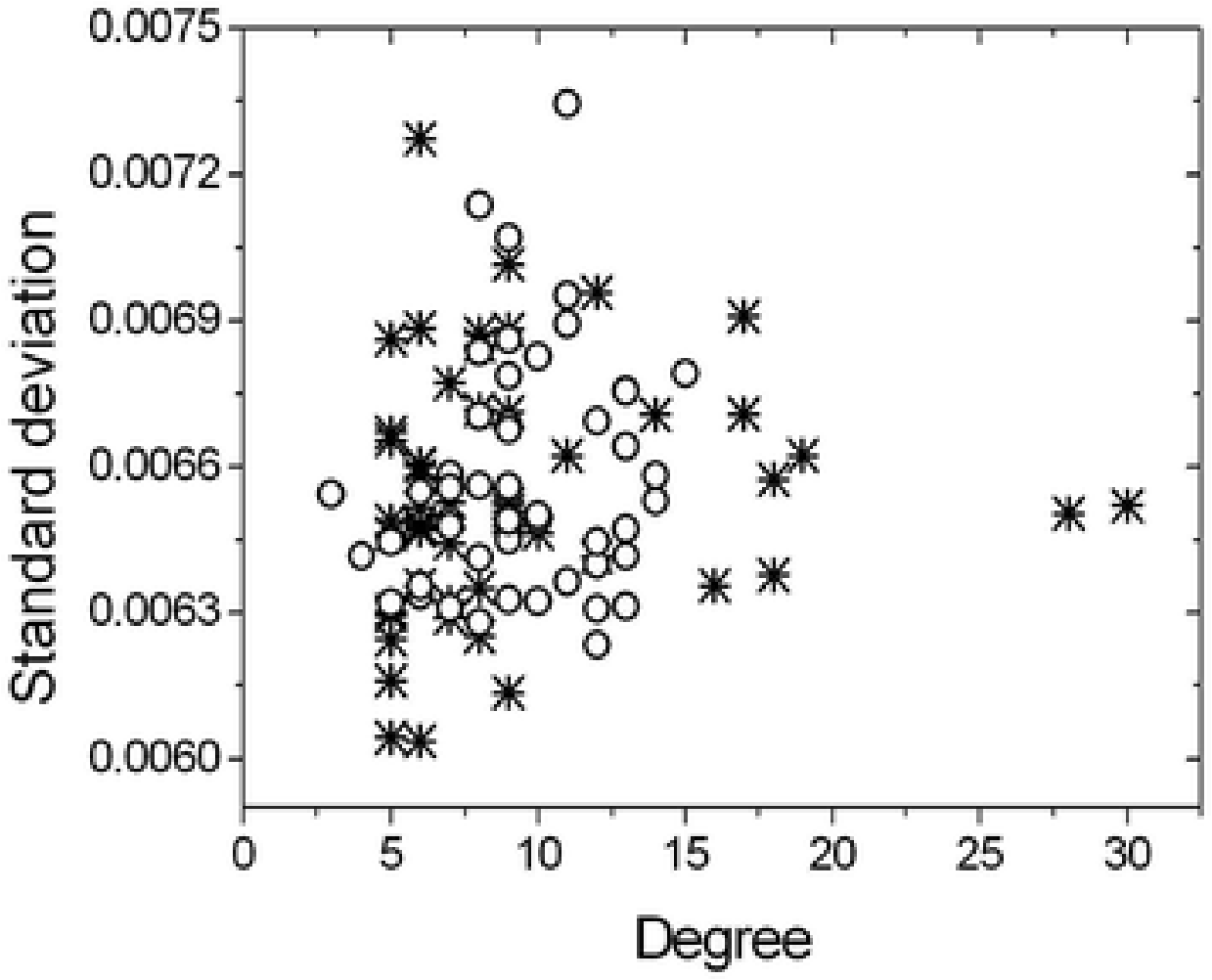}
  \includegraphics[width=0.3\linewidth]{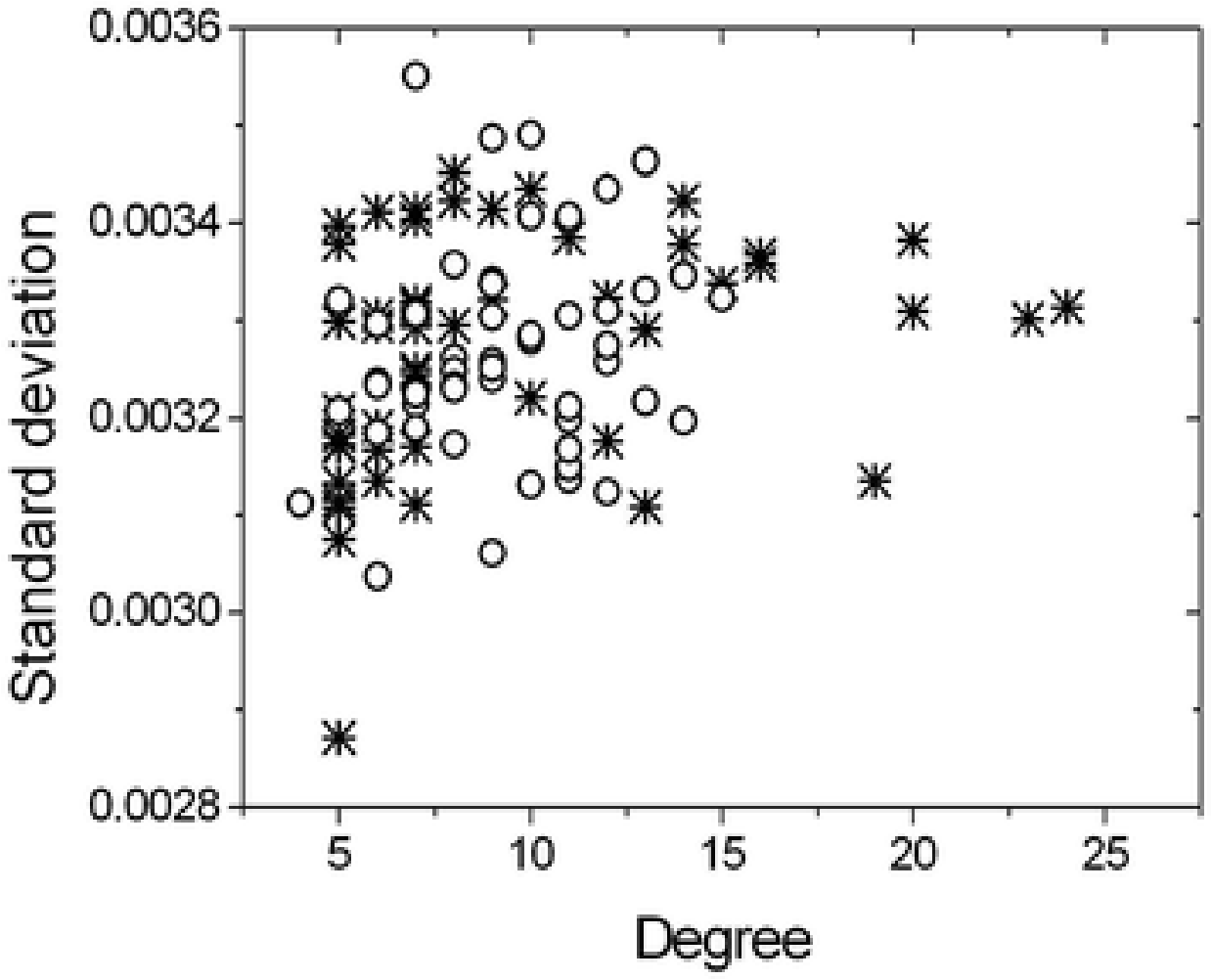}  
  \includegraphics[width=0.3\linewidth]{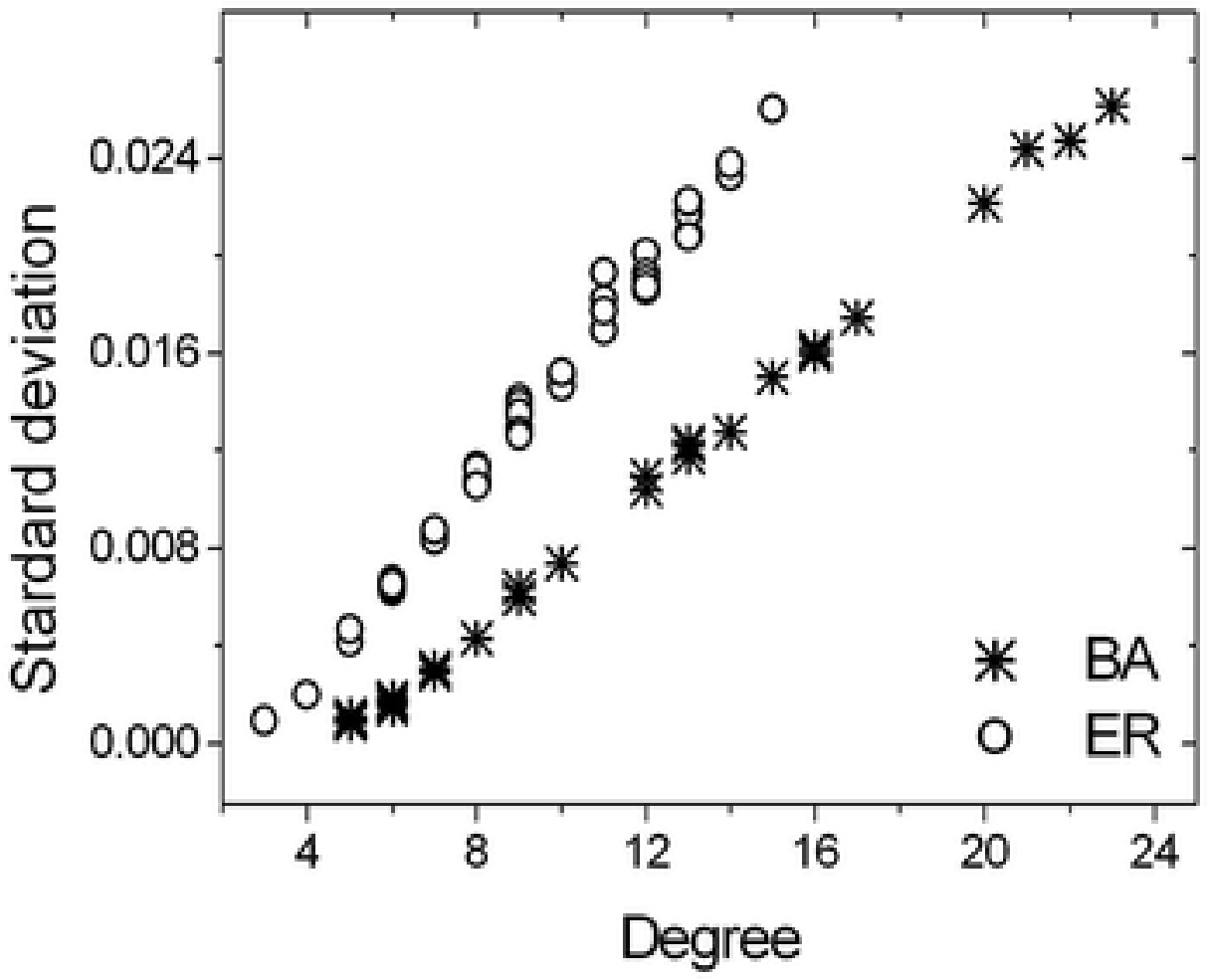} \\
  (a) \hspace{5cm} (b) \hspace{5cm} (c) \\ \vspace{0.5cm} 
  \caption{The standard deviation of individual errors in terms of the
  degree of the agents with different error rates: (a) one agent with
  $\gamma = 0.2$ and all other agents with 0.4; (b) one node with
  $\gamma= 0.4$ and all other nodes with 0.2; and (c) one agent with
  an error proportional to its degree and all other agents with
  0.2. }~\label{fig:stdev} \end{center}
\end{figure*}

Figure~\ref{fig:stdev} shows the standard deviation of individual
errors in terms of the degrees of the agents with different error rate
for the three situations above. Differently from the first two
situations (figures ~\ref{fig:stdev}(a) and ~\ref{fig:stdev}(b)), we
detected a positive correlation when one agent has an
error rate proportional to its degree (figure ~\ref{fig:stdev}(c)).

\subsection{Varying Observation Errors in Communities}

\begin{figure}[htb]
  \vspace{0.3cm} 
  \begin{center}
  \includegraphics[width=0.9\linewidth]{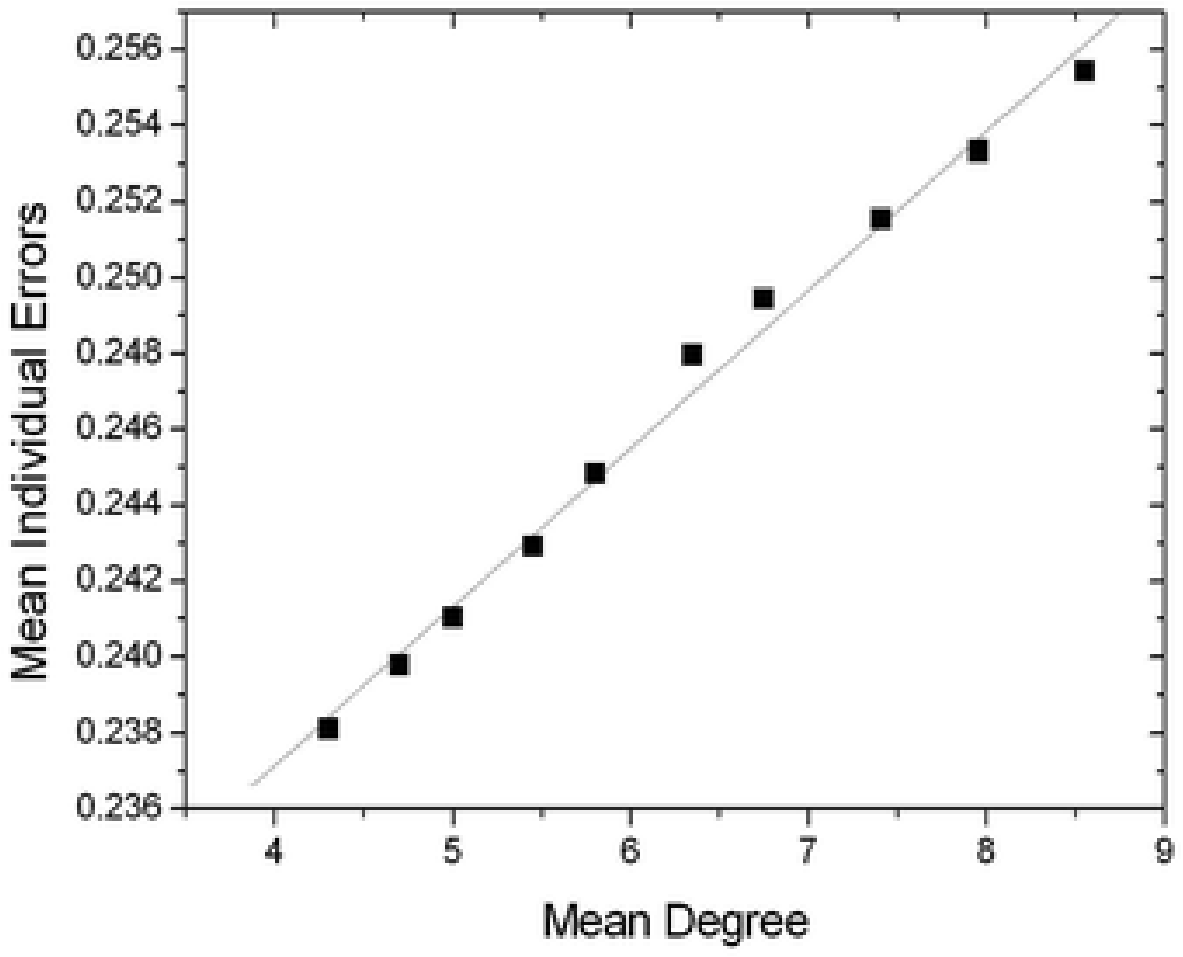} \\
  \vspace{0.5cm}
  \caption{Correlation between the average degree of the community of
  higher error rate and the mean value of the individual errors of the
  agents of the other communities. }~\label{fig:comm} \end{center}
\end{figure}

We also applied the knowledge acquisition dynamics to a network of
agents with community structure. During the simulations all agents of
one of the communities have twice the error rate of the others. We are
interested in obtaining the correlations between the average degree of
the community with higher error rate and the mean value of the
individual errors of the agents of the other communities.
Figure~\ref{fig:comm} shows the results for a network with $N_a=100$
agents and 5 communities of the same size.  While the estimation error
of one of the communities is 0.4, the agents of other communities have
error 0.2. We find a positive correlation which means that the
addition of edges between the agents with higher error rates increases
the individual errors of the agents of the other communities.

This effect can be explained by the existence of bridging agents which
belong to the community with higher error rate but connect to another
clusters. Because the individual errors takes into account the models
of the neighborhood, these special agents tend to reach worse models
when the intra-community communication is intensified. Indeed, the
addition of new edges implies more agents with higher errors to
influence to the development of their estimations. Consequently, the
loss of performance propagates to another communities.

\section{Concluding Remarks}

The important problem of scientific interaction has been effectively
investigated in terms of concepts and tools of complex networks
research (e.g.\cite{Price65:Science, Newman01_c,
Newman01_a,Newman01_b, Moody04, Barabasi02}).  However, most of the
interest has been so far concentrated on characterizing the
connectivity between scientists and institutions. The present study
reported what is possibly the first approach to model the dynamics of
knowledge acquisition (building a model of a complex system) by a
system of multiple agents interacting through specific types of
complex network. Several configurations were considered at increasing
levels of sophistication. Each agent was assumed to make an
observation of the system of interest at each time step with error
probability $\gamma$. 

A series of interesting and important results have been identified
analytically and through simulations. First, we have that the
individual estimation error tends to $\gamma$ when the agents do not
interact one another. However, different individual errors were
observed when the agents were allowed to consider the average of
models at each of their immediate neighbors. Special attention was
given to the cases in which one of the agents has a different
observation error, yielding the important result that the overall
error in such a configuration tends to be correlated with the degree
of the agent with different observation error. More specifically, we
demonstrated that this correlation is linear when one agent has an
error half or twice the error rate of the others, and nonlinear when
it is proportional to its degree. In other words, the connectivity
will have a substantial influence over the models developed by the
agents, for better or for worse. It is interesting to observe that
agents with many connections will imply strong influences over the
whole network even in case those agents have no special fitness. Such
an effect is a direct consequence of the fact that those agents are
heard by more people. In case the hubs have higher observation errors,
worse models are obtained throughout the agents network. In
particular, the negative influence of the hubs is more strongly felt
by its neighborhood. Finally, we have shown that when the agents are
clustered into communities, the addition of edges in the communities
with different error rate leads to the change of the individual errors
of the agents in the other groups.

This investigation has paved the way to a number of subsequent works,
including but not being limited to: consideration of model degradation
along time, other learning strategies, other types of networks,
observation errors conditional to specific local features (e.g. degree
or clustering coefficient) of the network being modeled, as well as
other distribution of observation errors among the agents.

\begin{acknowledgments}
Luciano da F. Costa thanks CNPq (308231/03-1) and FAPESP (05/00587-5)
for sponsorship.  Part of this work was developed during the author's
Visiting Scholarship at St. Catharine's College, University of
Cambridge, UK.
\end{acknowledgments}

\bibliography{netmodel}
\end{document}